\documentclass[pop,fleqn]{w-art}
\usepackage{amssymb}
\usepackage{graphicx}
\newcommand\rf[1]{(\ref{eq:#1})}
\newcommand\lab[1]{\label{eq:#1}}
\newcommand\nonu{\nonumber}
\newcommand\br{\begin{eqnarray}}
\newcommand\er{\end{eqnarray}}
\newcommand\be{\begin{equation}}
\newcommand\ee{\end{equation}}

\newcommand\lb{\lbrack}
\newcommand\rb{\rbrack}

\newcommand\llb{\left\lbrack}
\newcommand\rrb{\right\rbrack}

\renewcommand\({\left(}
\renewcommand\){\right)}
\newcommand\bv{\bigm\vert}               %%
\newcommand\bgv{\bigg\vert}              %%

\newcommand\bc{\begin{center}}
\newcommand\ec{\end{center}}

\newcommand\partder[2]{\frac{{\partial {#1}}}{{\partial {#2}}}}
\renewcommand\b{\beta}

\renewcommand\d{\delta}

\newcommand\vareps{\varepsilon}
\newcommand\g{\gamma}
\newcommand\G{\Gamma}

\newcommand\h{\frac{1}{2}}
\renewcommand\k{\kappa}
\renewcommand\l{\lambda}

\newcommand\m{\mu}
\newcommand\n{\nu}

\newcommand\p{\phi}
\newcommand\vp{\varphi}
\renewcommand\P{\Phi}
\newcommand\pa{\partial}

\newcommand\pr{\prime}

\renewcommand\r{\rho}
\newcommand\s{\sigma}

\renewcommand\t{\tau}
\renewcommand\th{\theta}

\newcommand\wti{\widetilde}
\newcommand\cA{{\mathcal A}}

\newcommand\cF{{\mathcal F}}

\newcommand{\ct}[1]{\cite{#1}}
\newcommand{\bib}[1]{\bibitem{#1}}
\newcommand\NPB[3]{\textsl{Nucl. Phys.} \textbf{B#1}, #3 (#2)}

\newcommand\PRD[3]{\textsl{Phys. Rev.} \textbf{D#1}, #3 (#2)}

\newcommand\PLB[3]{\textsl{Phys. Lett.} \textbf{#1B}, #3 (#2)}
\newcommand\CQG[3]{\textsl{Class. Quantum Grav.} \textbf{#1}, #3 (#2)}

\newcommand\IJMPA[3]{\textsl{Int. J. Mod. Phys.} \textbf{A#1}, #3 (#2)}

\newcommand\MPLA[3]{\textsl{Mod. Phys. Lett.} \textbf{A#1}, #3 (#2)}

\begin{document}

%%    The information for the title page will be placed between
%%    \begin{document} and \maketitle. The order of most entries
%%    is determined by the class file and can not be changed by
%%    rearranging them. The maketitle command follows after the
%%    abstract.
%%
%%    Most of the following commands will be completed by the publisher.
%%
%%    The copyrightyear is defined in the .clo file as the first argument
%%    of the copyrightinfo command. If the copyrightyear differs from that
%%    value it might be adjusted by the following definition:
%%
%% \renewcommand{\copyrightyear}{2003}% uncomment to change the copyrightyear.
%%
\DOIsuffix{theDOIsuffix}
%%
%% issueinfo for header and copyright line
\Volume{}
\Issue{}
\Month{}
\Year{}
%%
%%    First and last pagenumber of the article. If the option
%%    'autolastpage' is set (default) the second argument may be left empty.
\pagespan{3}{}
%%
%%    Dates will be filled in by the publisher. The 'reviseddate' and
%%    'dateposted' (Published online) entry may be left empty.
\Receiveddate{}
\Reviseddate{}
\Accepteddate{}
\Dateposted{}
\keywords{Weyl-conformal invariance, lightlike branes, black holes.}
\subjclass[pacs]{11.25.-w, 04.70.-s, 04.50.+h}

%% \pretitle{Editor's Choice}

%% We have a short and a long form for the title. The short form
%% (optional argument) goes into the running head.

\title[Lightlike Branes]{Weyl-Invariant Lightlike Branes and Soldering of Black
Hole Space-Times}

%% Please do not enter footnotes or \inst{}-notes into the optional
%% argument of the author command. The optional argument will go into
%% the header.  If there is only one address the marker \inst{x} may be
%% omitted.

%% Information for the first author.
\author[F. Author]{E.I. Guendelman\inst{1,}\footnote{E-mail: 
~\textsf{guendel@bgumail.bgu.ac.il}.}}
\address[\inst{1}]{Department of Physics, Ben-Gurion University of the Negev,
P.O.Box 653\\ IL-84105 ~Beer-Sheva, Israel}
%%
%%    Information for the second author
\author[S. Author]{A. Kaganovich\inst{1,}\footnote{E-mail: 
~\textsf{alexk@bgumail.bgu.ac.il}.}}
%%
%%    Information for the third author
\author[T. Author]{E. Nissimov\inst{2,}
  \footnote{Corresponding author\quad E-mail:~\textsf{nissimov@inrne.bas.bg},
            Phone: +359\,2\,7144\,720,
            Fax: +359\,2\,975\,3619}}
\address[\inst{2}]{Institute for Nuclear Research and Nuclear Energy,
Bulgarian Academy of Sciences\\
Boul. Tsarigradsko Chausee 72, BG-1784 ~Sofia, Bulgaria}
%%
%%    Information for the fourth author
\author[]{S. Pacheva\inst{2,}\footnote{E-mail: ~\textsf{svetlana@inrne.bas.bg}.}}
%%
%%    \dedicatory{This is a dedicatory.}

\begin{abstract}
We consider self-consistent coupling of the recently introduced new class
of Weyl-conformally invariant {\em lightlike} branes (\textsl{WILL}-branes)
to $D\! =\! 4$ Einstein-Maxwell system plus a $D\! =\! 4$ three-index antisymmetric 
tensor gauge field. We find static spherically-symmetric solutions
where the space-time consists of two regions with different black-hole-type 
geometries and different values for a {\em dynamically generated} cosmological
constant, separated by the \textsl{WILL}-brane which ``straddles'' their common
event horizon. Furthermore, the \textsl{WILL}-brane produces a potential ``well''
around itself acting as a trap for test particles falling towards the horizon.
\end{abstract}

\maketitle

%% If there is not enough space inside the running head
%% for all authors including the title you may provide
%% the leftmark in one of the following three forms:

%% \renewcommand{\leftmark}
%% {First Author: A Short Title}

%% \renewcommand{\leftmark}
%% {First Author and Second Author: A Short Title}

\renewcommand{\leftmark}
{E.I. Guendelman et al.: Lightlike Branes}

%% \tableofcontents  % Produces the table of contents.

%%%%%%%%%%%%%%%%%%%%%%%%%%%%%%%%%%%%%%%%%%%%%%%%%%%%%%%%%%%%%%%%%%%%%%%%%%%%%%%%
\section{\label{sec:intro}Introduction}

Lightlike membranes are of particular interest in general relativity as they
describe impulsive lightlike signals arising in various violent astrophysical
events, \textsl{e.g.}, final explosion in cataclysmic processes such as supernovae
and collision of neutron stars \ct{barrabes-hogan}. 
% In the context of gravity and cosmology
Lightlike membranes are basic ingredients in the so called ``membrane paradigm'' 
theory \ct{membrane-paradigm} which appears to be a quite effective treatment of the
physics of a black hole horizon. 

In refs.\ct{Israel-66,Barrabes-Israel-Hooft} lightlike membranes in the context of
gravity and cosmology have been extensively studied from a phenomenological point
of view, \textsl{i.e.}, by introducing them without specifying the Lagrangian 
dynamics from which they may originate. Recently in a series of papers
\ct{will-brane-kiten-zlatibor,will-prd-trap} we have developed a new field-theoretic 
approach for a systematic description of the dynamics of lightlike branes starting 
from concise {\em Weyl-conformally invariant} actions. The latter are 
related to, but bear significant qualitative differences from, the standard 
Nambu-Goto-type $p$-brane actions\footnote{In ref.\ct{barrabes-israel-05} brane actions 
in terms of their pertinent extrinsic geometry have been proposed which generically
describe non-lightlike branes, whereas the lightlike branes are treated as a limiting
case.} (here $(p+1)$ is the dimension of the brane world-volume).

In the present note we discuss spherically-symmetric solutions for the coupled 
system of bulk $D\! =\! 4$ Einstein-Maxwell plus $3$-index antisymmetric tensor
gauge field interacting with a \textsl{WILL}-brane. The latter serves as a 
matter and charged source for gravity and electromagnetism and, in addition, 
produces a space-varying dynamical cosmological constant. The above solutions
describe space-times divided into two separate regions with different black hole 
geometries and different values of the dynamically generated cosmological
constant, separated by the \textsl{WILL}-brane which automatically position
itself on (``straddles'') their common horizon. The matching of the physical 
parameters of the two black hole space-time regions (``soldering'') is 
explicitly given in terms of the free \textsl{WILL}-brane coupling parameters
(electric surface charge density and Kalb-Rammond coupling constant).
A physically intersting implication of the above solutions is the emergence of
a potential ``well'' around the \textsl{WILL}-brane trapping infalling test
particles towards the common horizon.

%%%%%%%%%%%%%%%%%%%%%%%%%%%%%%%%%%%%%%%%%%%%%%%%%%%%%%%%%%%%%%%%%%%%%%%%%%%%%%%%
\section{\label{sec:will-brane}Weyl-Conformally Invariant Lightlike Branes}

In refs.\ct{will-brane-kiten-zlatibor,will-prd-trap} we proposed the following
new kind of $p$-brane action (in what follows we shall concentrate on the
first nontrivial case $p\! =\! 2$) :
\br
S = - \int d^3\s \,{\P (\vp)}
\Bigl\lb \h \g^{ab} \pa_a X^{\m} \pa_b X^{\n} G_{\m\n}
- {\sqrt{F_{ab} F_{cd} \g^{ac}\g^{bd}}}\Bigr\rb
\nonu \\
-\; q\int d^3\s \,\vareps^{abc} \cA_\m \pa_a X^\m F_{bc} \;
{
- \;\frac{\b}{3!} \int d^3\s \,\vareps^{abc} \pa_a X^\m \pa_b X^\n \pa_c X^\l \cA_{\m\n\l}}
\lab{WILL-brane+A+A3}
\er
% As usual $\g_{ab}$ denotes the intrinsic world-volume Riemannian metric.
The first significant difference of \rf{WILL-brane+A+A3} w.r.t. standard
Nambu-Goto-type $p$-brane action is the presence of a new non-Riemannian 
reparametrization-covariant integration measure density: %(volume form)
$\P (\vp) \equiv \frac{1}{3!} \vareps_{ijk}
\vareps^{abc} \pa_a \vp^i \pa_b \vp^j \pa_c \vp^k \;,\;
(a,b,c =0,1,2\; ,\; i,j,k=1,2,3)$,
built in terms of auxiliary world-volume scalar fields $\vp^i$. 
As usual $\g_{ab}$ denotes the intrinsic Riemannian metric on the brane
world-volume and $\g \equiv \det\Vert\g_{ab}\Vert$. The second important
difference is the ``square-root'' Maxwell term\footnote{``Square-root'' 
Maxwell (Yang-Mills) action in
$D\! =\! 4$ was originally introduced in the first ref.\ct{Spallucci}
and later generalized to ``square-root'' actions of higher-rank
antisymmetric tensor gauge fields in $D\geq 4$ in the second and
third refs.\ct{Spallucci}.} involving an auxiliary 
world-volume gauge field $A_a$ with $F_{ab} = \pa_a A_b - \pa_b A_a$.
$G_{\m\n}$ ($\m,\n = 0,1,2,3$) denotes Riemannian metric on the embedding
$D\! =\! 4$ space-time.
The second Chern-Simmons-like term in \rf{WILL-brane+A+A3}, describing a 
coupling to {external $D\! =\! 4$ space-time electromagnetic field $\cA_\m$}, 
is a special case of a class of Chern-Simmons-like couplings of extended objects
to external electromagnetic fields proposed in ref.\ct{Aaron-Eduardo}. 
The last term is a Kalb-Ramond-type coupling to external space-time rank 3 
gauge potential $\cA_{\m\n\l}$.

The action \rf{WILL-brane+A+A3} is manifestly invariant under {Weyl (conformal)
symmetry}: $\g_{ab}\!\! \longrightarrow\!\! \g^{\pr}_{ab} = \rho\,\g_{ab}$,
$\vp^{i} \longrightarrow \vp^{\pr\, i} = \vp^{\pr\, i} (\vp)$ with Jacobian 
$\det \Bigl\Vert \frac{\pa\vp^{\pr\, i}}{\pa\vp^j} \Bigr\Vert = \rho$.

Let us recall the physical significance of $\cA_{\m\n\l}$ \ct{Aurilia-Townsend}. 
In $D=4$ when adding kinetic term for $\cA_{\m\n\l}$ coupled to gravity
(see Eq.\rf{E-M-WILL} below), its field-strength
~$\cF_{\k\l\m\n} = 4\pa_{[\k}\cA_{\l\m\n]} = \cF \sqrt{-G} \vareps_{\k\l\m\n}$
with a single independent component $\cF$ produces {\em dynamical (positive)
cosmological constant} ~$K = \frac{4}{3}\pi G_N \cF^2$.

Invariance under world-volume reparametrizations allows to introduce the
standard (synchronous) gauge-fixing conditions:
$\g^{0i} = 0 \; (i=1,2) \;,\; \g^{00} = -1$. With the latter gauge choice
and using the short-hand notation 
$\(\pa_a X \pa_b X\)\equiv \pa_a X^\m G_{\m\n} \pa_b X^\n$, 
the equations of motion for the brane action \rf{WILL-brane+A+A3} read:
\br
{\(\pa_0 X \pa_0 X\) = 0 \quad ,\quad \(\pa_0 X \pa_i X\) = 0}  \quad ,\quad
{\(\pa_i X\pa_j X\) - \h \g_{ij} \g^{kl}\(\pa_k X\pa_l X\) = 0} \; ,
\lab{constr-0-vir}
\er
these are in fact constraints analogous to the (classical) Virasoro constraints
of string theory;
% (the latter look exactly like the classical (Virasoro) constraints for an
% Euclidean string theory w.r.t. $(\s^1,\s^2)$);
\br
\pa_i X^\m \pa_j X^\n \cF_{\m\n}(\cA) = 0 \quad ,\quad
\pa_i \chi + \sqrt{2} q \pa_0 X^\m \pa_i X^\n \cF_{\m\n}(\cA) = 0 \; ,
\lab{A-eqs-1}
\er
(here $\chi \equiv \frac{\P (\vp)}{\sqrt{-\g}}$ plays the role of {\em variable
brane tension}, $\cF_{\m\n} (\cA) = \pa_\m \cA_\n - \pa_\n \cA_\m$);
\br
{\wti{\Box}}^{(3)} X^\m + \( - \pa_0 X^\n \pa_0 X^\l +
\g^{kl} \pa_k X^\n \pa_l X^\l \) \G^{\m}_{\n\l}
\nonu \\
-\; q \frac{\g^{kl}\(\pa_k X \pa_l X\)}{\sqrt{2}\,\chi}\,
% \pa_0 X^\n \(\pa_\l \cA_\n - \pa_\n \cA_\l\)\, G^{\l\m} = 0
\pa_0 X^\n \cF_{\l\n}\, G^{\l\m}
% \nonu \\
- \frac{\b}{3!} \vareps^{abc} \pa_a X^\k \pa_b X^\l \pa_c X^\n G^{\m\r}\,
\cF_{\r\k\l\n} = 0 \; ,
\lab{X-eqs-1}
\er
where $\cF_{\r\k\l\n} = 4\pa_{[\k}\cA_{\l\m\n]}$ as above,
% $\cF_{\r\k\l\n} = 4 \pa_{[\r} \cA_{\k\l\n]}$, and
${{\wti \Box}^{(3)} \equiv
- \frac{1}{\chi \sqrt{\g^{(2)}}} \pa_0 \(\chi \sqrt{\g^{(2)}} \pa_0 \) +
\frac{1}{\chi \sqrt{\g^{(2)}}}\pa_i \(\chi \sqrt{\g^{(2)}} \g^{ij} \pa_j \)}$,
where $\g^{(2)} \equiv \det\Vert\g_{ij}\Vert$ ($i,j=1,2$),
and $\G^\m_{\n\l}=
\h G^{\m\k}\(\pa_\n G_{\k\l}+\pa_\l G_{\k\n}-\pa_\k G_{\n\l}\)$
is the affine connection corresponding to the external space-time
metric $G_{\m\n}$.

The first Virasoro-like constraint in \rf{constr-0-vir} explicitly exhibits the
inherent lightlike property of the brane model \rf{WILL-brane+A+A3}, 
hence the acronym \textsl{WILL} (Weyl-invariant light-like) brane.

\section{\label{sec:bulk-will-brane}Bulk Gravity-Matter Coupled to {\em WILL}-brane}

Let us now consider the coupled Einstein-Maxwell-\textsl{WILL}-brane 
system adding also a coupling to a rank 3 gauge potential:
\br
{S = \int\!\! d^4 x\,\sqrt{-G}\,\llb \frac{R(G)}{16\pi G_N}
% - \frac{1}{4} \cF_{\m\n} \cF_{\k\l} G^{\m\k} G^{\n\l}
- \frac{1}{4} \cF_{\m\n}\cF^{\m\n} - \frac{1}{4! 2 } \cF_{\k\l\m\n}\cF^{\k\l\m\n}\rrb
+ S_{\mathrm{WILL-brane}}} \; .
\lab{E-M-WILL}
\er
Here $\cF_{\m\n} = \pa_\m \cA_\n - \pa_\n \cA_\m$,
$\cF_{\k\l\m\n} = 4 \pa_{[\k} \cA_{\l\m\n]} = \cF \sqrt{-G} \vareps_{\k\l\m\n}$
as above, and the \textsl{WILL}-brane action is the same as in 
\rf{WILL-brane+A+A3}.

The equations of motion for the \textsl{WILL}-brane subsystem are the same as
\rf{constr-0-vir}--\rf{X-eqs-1}, whereas the equations for the space-time 
fields read:
\br
R_{\m\n} - \h G_{\m\n} R =
8\pi G_N \( T^{(EM)}_{\m\n} + T^{(rank-3)}_{\m\n} + T^{(brane)}_{\m\n}\) \; ,
\lab{Einstein-eqs}  \\
\pa_\n \(\sqrt{-G}G^{\m\k}G^{\n\l} \cF_{\k\l}\) + 
q \int\!\! d^3 \s\,\d^{(4)}\Bigl(x-X(\s)\Bigr)
\vareps^{abc} F_{bc} \pa_a X^\m = 0 \; ,
\lab{Maxwell-eqs}  \\
\vareps^{\l\m\n\k} \pa_\k \cF + \b\, \int\! d^3\s\, \d^{(4)}(x - X(\s))
\vareps^{abc} \pa_a X^{\l} \pa_a X^{\m} \pa_a X^{\n} = 0 \; .
\lab{F4-eqs}
\er
% where in the last equation we have used relation \rf{F4}. 
% The explicit form of the energy-momentum tensors read:
% \br
% T^{(EM)}_{\m\n} = \cF_{\m\k}\cF_{\n\l} G^{\k\l} - G_{\m\n}\frac{1}{4}
% \cF_{\r\k}\cF_{\s\l} G^{\r\s}G^{\k\l} \; ,
% \lab{T-EM} \\
The energy-momentum tensors read:
$T^{(EM)}_{\m\n} = \cF_{\m\k}\cF_{\n\l} G^{\k\l} - G_{\m\n}\frac{1}{4}
\cF_{\r\k}\cF_{\s\l} G^{\r\s}G^{\k\l}$,
\br
T^{(rank-3)}_{\m\n} = \frac{1}{3!}\llb \cF_{\m\k\l\r} {\cF_{\n}}^{\k\l\r} -
\frac{1}{8} G_{\m\n} \cF_{\k\l\r\s} \cF^{\k\l\r\s}\rrb = - \h \cF^2 G_{\m\n} \; ,
\lab{T-rank3} \\
T^{(brane)}_{\m\n} = - G_{\m\k}G_{\n\l}
\int\!\! d^3 \s\, \frac{\d^{(4)}\Bigl(x-X(\s)\Bigr)}{\sqrt{-G}}\,
\chi\,\sqrt{-\g} \g^{ab}\pa_a X^\k \pa_b X^\l  \; .
% \chi (\g,u,\cA) \sqrt{-\g} \g^{ab}\pa_a X^\k \pa_b X^\l
\lab{T-brane}
\er
% (recall $\chi\equiv\frac{\P(\vp)}{\sqrt{-\g}}$ -- the variable brane tension).

For the bulk gravity-matter system coupled to a charged \textsl{WILL}-brane 
\rf{E-M-WILL} we find the following static spherically symmetric solutions.
The bulk space-time consists of two regions separated by the \textsl{WILL}-brane
sitting on (``straddling'') a common horizon of the former:
\br
{(ds)^2 = - A_{(\mp)}(r)(dt)^2 + \frac{1}{A_{(\mp)}(r)}(dr)^2 +
r^2 \lb (d\th)^2 + \sin^2 (\th)\,(d\p)^2\rb} \; ,
\lab{2-regions}
\er
where the subscript $(-)$ refers to the region inside, whereas the subscript 
$(+)$ refers to the region outside the horizon at 
$r=r_0 \equiv r_{\mathrm{horizon}}$ with $A_{(\mp)}(r_0)=0$. The interior
region is a Schwarzschild-de-Sitter space-time:
\br
A(r)\equiv A_{(-)}(r) = 1 - K_{(-)} r^2 - \frac{2G_N M_{(-)}}{r}
\;\; ,\quad \mathrm{for}\;\; r < r_0 \; ,
\lab{schwarzschild-dS-inside}
\er
whereas the exterior region is  Reissner-Norstr\"{o}m-de-Sitter space-time:
\br
{A(r)\equiv A_{(+)}(r) = 1 - K_{(+)} r^2 - \frac{2G_N M_{(+)}}{r} +
\frac{G_N Q^2}{r^2}\;\; , \quad \mathrm{for}\;\; r > r_0} \; ,
\lab{RN-dS-outside}
\er
with Reissner-Norstr\"{o}m (squared) charge given by $Q^2 = 8\pi q^2 r_{0}^4$.
The rank 3 tensor gauge potential together with its Kalb-Rammond-type
coupling to the \textsl{WILL}-brane produce via Eq.\rf{F4-eqs} a dynamical 
space-varying cosmological constant which is different inside and outside the 
horizon:
$K_{(\pm)} = \frac{4}{3}\pi G_N \cF_{(\pm)}^2$ for $r \geq r_{0} \;\;
(\, r \leq r_{0}\,)$, $\cF_{(+)} = \cF_{(-)} - \b$.
The Einstein Eqs.\rf{Einstein-eqs} and the $X^\m$-brane Eqs.\rf{X-eqs-1}
yield two matching conditions for the normal derivatives w.r.t. the horizon
of the space-time metric components:
\br
\(\pa_r A_{(+)} - \pa_r A_{(-)}\)\!\!\bv_{r=r_0} = - 16\pi G_N \chi \;,\;
\(\pa_r A_{(+)} - \pa_r A_{(-)}\)\!\!\bv_{r=r_0} =
- \frac{r_0 (2q^2 + \b^2) \pa_r A_{(-)}\bv_{r=r_0} }{2\chi + \b r_0 \cF_{(-)}}
\;.
\nonu
\er
% \nonu \\
% \phantom{aaa}
% \lab{metric-matching}
% \er
The latter conditions allow to express all physical parameters of 
the solution, \textsl{i.e.}, two spherically symmetric black hole space-time
regions ``soldered'' along a common horizon via the \textsl{WILL}-brane in terms
of 3 free parameters $(q,\b,\cF)$ where (cf. Eq.\rf{WILL-brane+A+A3}):
(a) $q$ is the \textsl{WILL}-brane surface electric charge density;
(b) $\b$ is the \textsl{WILL}-brane (Kalb-Rammond-type) charge w.r.t. rank 3 
space-time gauge potential $\cA_{\l\m\n}$;
(c) $\cF_{(-)}$ is the vacuum expectation value of the 4-index field-strength 
$\cF_{\k\l\m\n}$ in the interior region. For the common horizon radius, the
Schwarzschild and Reissner-Nordstr\"{o}m masses we obtain:
\br
r_0^2 = \frac{1}{4\pi G_N \(\cF_{(-)}^2  - \b\cF_{(-)} + q^2 +\frac{\b^2}{2}\)}
\quad, \quad
M_{(-)} = \frac{r_0\,\(\frac{2}{3}\cF_{(-)}^2  - \b\cF_{(-)} + q^2 +\frac{\b^2}{2} \)}
{2G_N \(\cF_{(-)}^2  - \b\cF_{(-)} + q^2 +\frac{\b^2}{2}\)} \; ,
\lab{r-horizon-schwarzschild-mass}\\
M_{(+)} = M_{(-)} + \frac{r_0}{2G_N \(\cF_{(-)}^2  - \b\cF_{(-)} + q^2 +\frac{\b^2}{2}\)}
\( 2q^2 + \frac{2}{3}\b\cF_{(-)} - \frac{1}{3}\b^2\) \; .
\lab{RN-mass}
\er
For the brane tension we get accordingly:
$\chi = \frac{r_0}{2}\( q^2 + \frac{\b^2}{2} - 2\b\cF_{(-)}\)$.

Using expressions \rf{r-horizon-schwarzschild-mass}--\rf{RN-mass} we find
for the slopes of the metric coefficients $A_{(\pm)}(r)$ at $r=r_0$:
\br
\pa_r A_{(+)}\!\!\bv_{r=r_0} = - \pa_r A_{(-)}\!\!\bv_{r=r_0} \;\; ,\;\;
\pa_r A_{(-)}\!\!\bv_{r=r_0} = 8\pi G_N \chi = 
4\pi G_N r_0\, \( q^2 + \frac{\b^2}{2} - 2\b\cF_{(-)}\) \; .
\lab{A-slope-tension}
\er
In view of \rf{A-slope-tension} (and assuming for definiteness $\b >0$) we
conclude:

(i) In the area of parameter space 
$\cF_{(-)} > \frac{q^2 + \frac{\b^2}{2}}{2\b}$ (\textsl{i.e.}, when $\chi < 0$
-- negative brane tension) the common horizon is: (a) the de-Sitter horizon 
from the point of view of the interior Schwarzschild-de-Sitter geometry;
(b) it is the external Reissner-Nordstr\"{o}m horizon (the larger one) from 
the point of view of the exterior Reissner-Nordstr\"{o}m-de-Sitter geometry.

(ii) In the opposite area of parameter space 
$\cF_{(-)} < \frac{q^2 + \frac{\b^2}{2}}{2\b}$ (\textsl{i.e.}, when $\chi > 0$
-- positive brane tension) the common horizon is: (a) the Schwarzschild horizon
from the point of view of the interior Schwarzschild-de-Sitter geometry;
(b) it is the internal (the smaller one) Reissner-Nordstr\"{o}m 
horizon from the point of view of the exterior Reissner-Nordstr\"{o}m-de-Sitter
geometry.

%%%%%%%%%%%%%%%%%%%%%%%%%%%%%%%%%%%%%%%%%%%%%%%%%%%%%%%%%%%%%%%%%%%%%%%%%%%%%%%%
% \section{\label{sec:trapping}Trapping Potential Well Around Common Horizon}
%% Test particle dynamics
%% Graphics

Now let us consider planar motion of a (charged) test patricle with mass $m$ 
and electric charge $q_0$ in a gravitational background given by the solutions in
Section \ref{sec:bulk-will-brane}. Conservation of energy yields 
$\frac{E^2}{m^2} = {r^\pr}^2 + V^2_{eff}(r)$ 
~($E,\, J$ -- energy and orbital momentum of the test particle; prime indicates
proper-time derivative) with:
\br
{V^2_{eff}(r) = A_{(-)}(r) \( 1 + \frac{J^2}{m^2 r^2}\)
+ \frac{2Eq_0}{m^2}\sqrt{2}q r_0 -
\frac{q_0^2}{m^2} 2q^2 r_0^2 \qquad \( r \leq r_0 \)}
\nonu \\
{V^2_{eff}(r) = A_{(+)}(r) \( 1 + \frac{J^2}{m^2 r^2}\)
+ \frac{2Eq_0}{m^2} \frac{\sqrt{2}q r_0^2}{r} -
\frac{q_0^2}{m^2} \frac{2q^2 r_0^4}{r^2} \qquad \( r \geq r_0 \)}
\er
where $A_{(\mp)}$ are the same as in \rf{schwarzschild-dS-inside} and 
\rf{RN-dS-outside}. Taking into account \rf{A-slope-tension} we see that 
in the parameter interval 
$\cF_{(-)} \in \(\frac{q^2 + \frac{\b^2}{2}}{\b}, \infty\)$
the (squared) effective potential $V^2_{eff}(r)$ acquires a potential ``well''
in the vicinity of the \textsl{WILL}-brane (the common horizon) with a
minimum on the brane itself. 

In the simplest physically interesting case with $q=0,\, \cF_{(-)}=\b$ and 
$\b$ -- arbitrary, \textsl{i.e.}, matching of Schwarzschild-de-Sitter
interior (with dynamically generated cosmological constant) 
against pure Schwarzschild exterior (with {\em no} cosmological constant)
along the \textsl{WILL}-brane as their common horizon, 
the typical form of $V^2_{eff}(r)$ is graphically depicted in Fig.1.

% FIGURE-2: 04_Talk-Napoli_pic-2.pdf
\begin{figure}[b]
\includegraphics{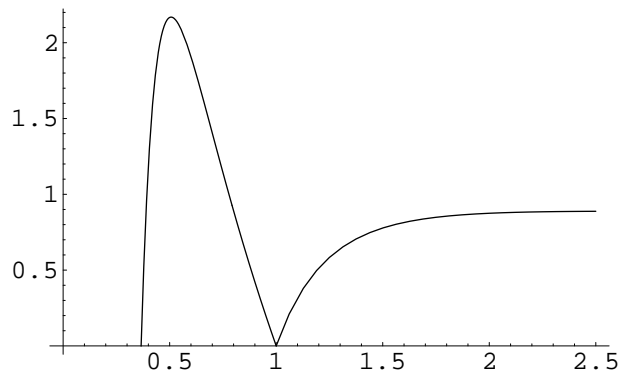}
\caption{Shape of $V^2_{eff}(r)$ as a function of the dimensionless ratio 
$x \equiv r/r_0$}
\end{figure}

Thus, we conclude that if a test particle moving towards the common event 
horizon loses energy (\textsl{e.g.}, by radiation), it may fall and be trapped 
by the potential well, so that it neither falls into the black hole nor can
escape back to infinity and, as a result, a ``cloud'' of trapped particles is 
formed around the \textsl{WILL}-brane materialized horizon.

%%%%%%%%%%%%%%%%%%%%%%%%%%%%%%%%%%%%%%%%%%%%%%%%%%%%%%%%%%%%%%%%%%%%%%%%%%%%%%%%
\textbf{Acknowledgements.}
% {\small Two of us (E.N. and S.P.) are sincerely grateful for hospitality and
% support to the organizers of the \textsl{Fourth Summer School on Modern
% Mathematical Physics} (Belgrade, Serbia, Sept. 2006), and the
% \textsl{Second Workshop of the European RTN} {\em ``Constituents,
% Fundamental Forces and Symmetries of the Universe''} (Naples, Italy, Oct. 2006), 
% where the above results were first presented. 
% {\em ``EUCLID''} (contract No.\textsl{HPRN-CT-2002-00325})

E.N. and S.P. are supported by European RTN network {\em ``Constituents, 
Fundamental Forces and Symmetries of the Universe''} (contract 
No.\textsl{MRTN-CT-2004-005104}).
They also received partial support from Bulgarian NSF grant \textsl{F-1412/04}.
Finally, all of us acknowledge support of our collaboration through the exchange
agreement between the Ben-Gurion University of the Negev (Beer-Sheva, Israel) and
the Bulgarian Academy of Sciences.
%%%%%%%%%%%%%%%%%%%%%%%%%%%%%%%%%%%%%%%%%%%%%%%%%%%%%%%%%%%%%%%%%%%%%%%%%%%%%%%%

\end{document}